\documentclass[12pt]{iopart}
\usepackage{graphicx}
\usepackage{orcidlink}
\usepackage{hyperref}

\begin{document}

\title[Trapped flux in small CaKFe$_4$As$_4$ crystal]{Trapped flux in a small crystal of CaKFe$_4$As$_4$ at ambient pressure and in a diamond anvil pressure cell}

\author{ Sergey L  Bud'ko\,\orcidlink{0000-0002-3603-5585}$^*$,\footnote[0]{$^*$Author to whom any correspondence should be addressed.} Shuyuan Huyan\,\orcidlink{0000-0003-0999-2440}, Mingyu Xu\orcidlink{0000-0001-7138-6009} and Paul C Canfield\orcidlink{0000-0002-7715-0643}}

\address{Ames National Laboratory and Department of Physics and Astronomy, Iowa State University, Ames, Iowa 50011, USA}
\ead{budko@ameslab.gov}
\vspace{10pt}

\begin{abstract}

In an extension of our previous work, [Sergey L Bud'ko et al 2023 {\it Supercond. Sci. Technol.} {\bf 36} 115001] the measurements of temperature dependent magnetization associated with trapped magnetic flux in a small single crystal of  CaKFe$_4$As$_4$,  using zero - field - cooled and field - cooled protocols  were performed, on the same crystal, at ambient pressure without a pressure cell and at 2.2 GPa in a commercial diamond anvil cell (DAC), showing comparable results. The data show that with a proper care and understanding, trapped flux measurements in superconductors indeed can be performed on samples in DACs under pressure, as was done on superhydrides [V S  Minkov et al  2023 {\it Nat. Phys.} {\bf 19} 1293]. 

\end{abstract}

%
%
\submitto{\SUST}
%
%
%

\section{Introduction}

The phenomenon of trapped flux in superconductors has been detected and examined for more than half a century. \cite{pip55a} In type II superconductors it is more pronounced and ubiquitous, with the general physical picture given by considering Bean's critical state model \cite{bea62a,bea64a} and the pinning of vortices. More recently, the interest in trapped flux in superconductors was shifted to potential applications (see e.g. Ref. \cite{hul10a}), however the importance of this phenomenon as one of the experimental proofs of superconductivity was well appreciated. \cite{mul87a} Indeed trapped flux measurements were used as one of the experimental confirmations of superconductivity in H$_3$S at high pressure. \cite{min23a} It was shown \cite{min23a}  that, in contrast to traditional dc magnetization measurements, the  trapped flux magnetization data are almost unaffected by the background signal of the diamond anvil cell due to the virtual absence of external magnetic fields in this measurement protocol. 

In Ref. \cite{bud23a} detailed measurements of magnetization associated with the flux trapped in superconducting crystals of MgB$_2$ as well as pure and Mn-substituted CaKFe$_4$As$_4$ at ambient pressure were reported. These data were proposed to serve as a baseline for interpretation of high pressure, trapped flux measurements in superhydrides. \cite{min23a} Indeed, there was a remarkable similarity between  the sets of data in Ref. \cite{min23a} and Ref. \cite{bud23a}. This said, though, the issue of sample size, and associated signal, could be considered a remaining point that still needed bench-marking.  The volume of our CaKFe$_4$As$_4$ sample in Ref. \cite{bud23a} was more than 1000 times larger that the sample volume that a high pressure DAC could be expected to  accommodate. 

In the desire to address concerns about sample size and signal associated with trapped flux measurements in a DAC,  we present trapped flux measurements on a small CaKFe$_4$As$_4$ sample both at ambient pressure, without a DAC and under 2.2 GPa pressure in a commercial DAC and compare  the resulting data sets with each other as well as with recent trapped flux result on hydride superconductors, also in DAC. \cite{min23a} We also briefly discuss possible modifications of measurements protocols and some challenges that the use of commercial DACs presents.

\section{Experimental details}

Single crystals of CaKFe$_4$As$_4$  with sharp superconducting transitions ($T_c = 34.7$~K) were grown using high-temperature solution growth. \cite{mei16a,mei17a} These crystals grow as mirrorlike, metallic, micaceous plates with the crystallographic $c$ axis is perpendicular to the plate surface (as determined by x-ray diffraction). The sample used in the measurements was a thin plate with the $c$ - axis perpendicular to the plate and approximately of a cuboid shape with some irregularities. The width and the length of the sample were measured using an optical microscope as $\sim 0.1 \times 0.08$~mm$^2$, whereas the thickness or the mass were not determined due to resolution of available instruments and a concern for keeping the sample intact. Instead, the mass of $\sim 0.5~\mu$g can be inferred from the low field magnetization measurements, shown in Fig. \ref{F1}, assuming complete shielding, and the density ($\sim 5.2$~g/cm$^3$) inferred from the unit cell parameters as determined by x-ray diffraction. \cite{iyo16a} The CaKFe$_4$As$_4$ crystal used in this work was cleaved and cut from one of the larger crystals, that showed no presence of secondary phases in magnetization measurements (see Ref. \cite{mei17a}  for the details of how the samples are pre-selected). The size of the sample was specifically chosen to approximately match the size of the H$_3$S sample measured in Ref. \cite{min23a}. As mentioned above, a detailed study of trapped flux in larger, mm-size, crystal of  CaKFe$_4$As$_4$ is presented in Ref. \cite{bud23a}.

Ambient pressure magnetization measurements were performed in a Quantum Design Magnetic Property Measurement System (MPMS3) SQUID magnetometer in the dc mode (30~mm scan length, a standard value for MPMS3) with a half-cylindrical quartz sample holder \cite{app} with a  small L – shaped  adapter (of the mass $\sim 4$~mg) made out of 0.05 mm thick copper foil  used to position the sample with $H \| c$. The sample was attached to the adapter by a thin layer of Dow Corning high vacuum grease. At ambient pressure we followed exactly the same measurements protocols as in Ref. \cite{bud23a}. In brief, in field-cooled (FC) protocol the sample was cooled down from above $T_c$ in the target field, $H_M$; after the temperature was stabilized at the target base temperature (1.8~K), the magnetic field was decreased to $H = 0$; then after 1 min dwell time, $M(T)$ measurements started. In zero-field cooled (ZFC) protocol: the sample was cooled down from above $T_c$ in $H = 0$; after the temperature was stabilized at the target base temperature, the magnetic field was increased to the target field, and then the magnetic field was decreased back to $H = 0$; then, after a 1 min dwell time, $M(T)$ measurements started. 

The DC magnetization measurements under high pressure were performed in a Quantum Design MPMS-classic SQUID magnetometer, with the standard for this unit scan length of 60~mm. (Note that the use of different models of MPMS instruments for measurements at ambient pressure and under pressure in DAC in this work was dictated by scheduling and availability reasons.) To ensure better thermalization of the DAC, for each temperature point, after instrument temperature stabilization a delay of 45~sec. was implemented. 300~sec, wait time was used after the change of magnetic field. The same wait / delay times were used  in our previous measurements with the same DAC in the same MPMS-classic (see e.g. \cite{huy24a}).  For more convenient temperature control, in the trapped flux measurements with DAC we used $T = 5$~K as a base temperature. A commercial DAC, (Almax - easyLab Mcell Ultra \cite{mcell}) with a pair of 700-$\mu$m-diameter culet-sized diamond anvils and tungsten gasket with 300-$\mu$m-diameter hole, was used. Nujol mineral oil, that solidifies at $\sim 1.3$~GPa at room temperature, served as the pressure medium.  \cite{ada82a} The pressure at room temperature was measured by the $R_1$ fluorescence line of a ruby ball. \cite{she20a} The background signal of the DAC with gasket, ruby ball and pressure medium but without sample was measured under 2~GPa using the same protocols as in the measurements with the sample. Given that our intent in this work is to use commercially available measurement systems as well as pressure cells, we do want to point out that there are ways to minimize the cell background even further by using specialty alloys rather than the more standard Be-Cu alloys.  In Appendix A we briefly address this point.

The magnetization of the sample was analyzed using a  point by point subtraction of the SQUID response with/without the sample, and then a dipole fitting on the resulting curve, following Ref. \cite{coa20a}. Examples of this procedure are shown in the Appendix B. Measurements under pressure were performed at one pressure point, 2.2~GPa. Although the chosen size of the sample  and the DAC would allow us to go to significantly higher pressure, we wanted to ensure that we are clearly below the $\sim 4$~GPa critical pressure of the structural transition of the CaKFe$_4$As$_4$ sample's transition to half collapsed tetragonal phase \cite{kal17a}, since no bulk superconductivity is observed above this critical pressure.

\section{Results}

\subsection{Ambient pressure}

\begin{figure}
\begin{center}
\includegraphics[angle=0,width=150mm]{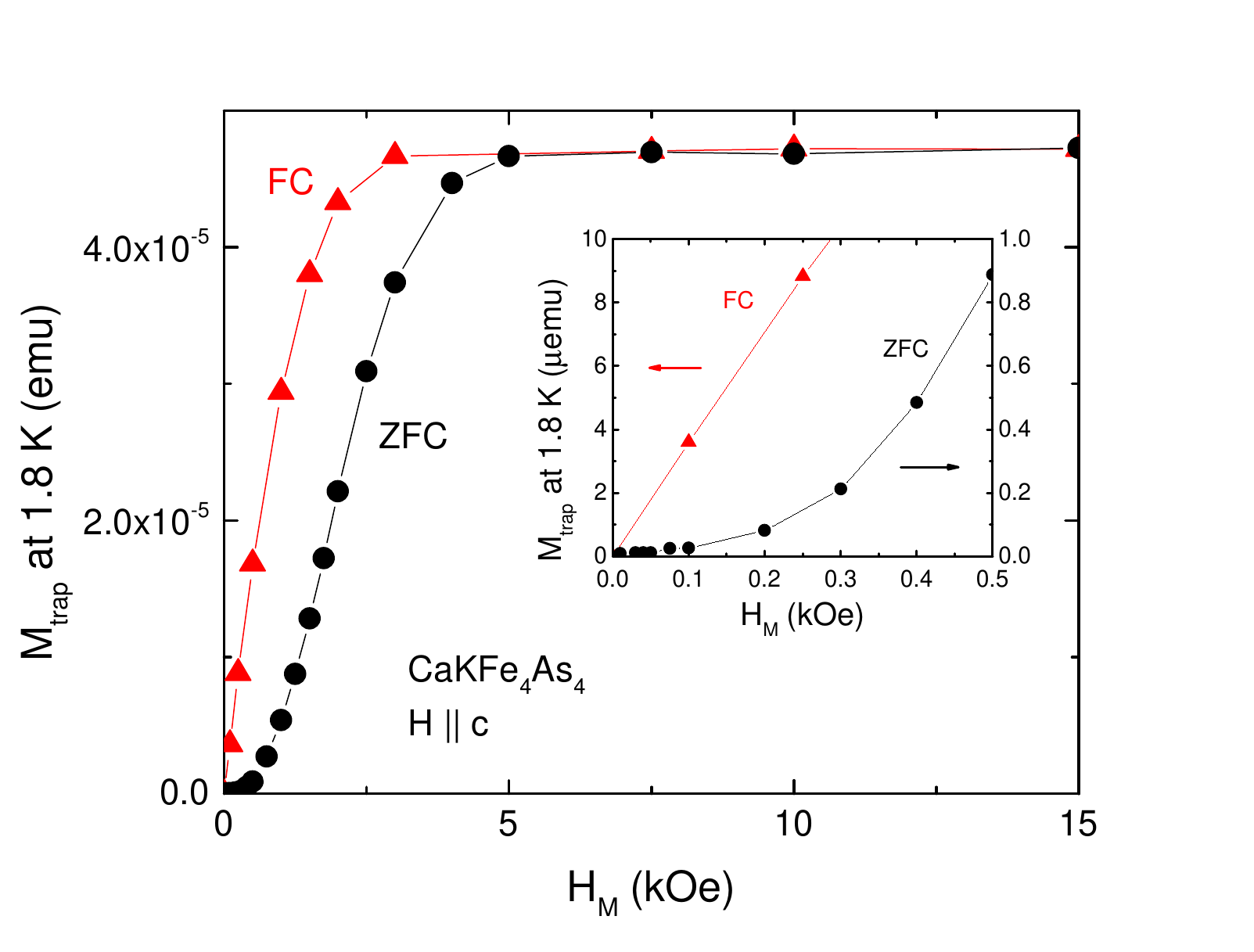}
\end{center}
\caption{(color online) Trapped field magnetization at $T = 1.8$~K as a function of target field $H_M$ in the ZFC and FC experiments. The inset shows an enlarged, low field, part of the data.} \label{F3} 
\end{figure}

Low field ZFC and FC temperature dependent magnetization of CaKFe$_4$As$_4$  is shown in Fig. \ref{F1}. The superconducting transition with $T_c \approx 34.7$~K is clearly seen in ZFC data, in agreement with the literature. \cite{bud23a,mei16a} No feature at $T_c$, within resolution, is observed in the FC data. As discussed in Refs. \cite{mei16a} and \cite{bud23a}, this is consistent with CaKFe$_4$As$_4$ having relatively large pinning. Note that the measured diamagnetic signal from the sample is comparable (albeit factor of $\sim 3$ larger, with the difference coming from the sample volume and demagnetization factor) to those reported  for H$_3$S and LaH$_{10}$. \cite{min22a} Data for the temperature-dependent magnetization associated with the trapped field taken using FC and ZFC protocols are shown in Fig. \ref{F2} and the summary plot of the trapped field magnetization at 1.8~K as a function of the target magnetic field $H_M$ using both FC and ZFC protocols is shown in Fig. \ref{F3}. Fig. \ref{F3} can be understood based on Bean's critical state model \cite{bea62a,bea64a} used for systems with pinning. The process of flux trapping is discussed in more details in Refs. \cite{min23a,mos91a,alt91a}. Taking differences in relative dimensions and demagnetization factors into account, these data are consistent with the previous study on a single crystal that was more than three orders of magnitude larger. \cite{bud23a} It should be noted that whereas the ZFC trapped flux data does go to zero at a finite field value, the FC trapped flux data goes through (0, 0) as any simple physical analysis would suggest (details of such analysis are provided in Ref. \cite{bud24a}).  This is in contrast to what is suggested in the empirical analysis presented by Hirsch and Marsiglio, \cite{hir22a} where both ZFC and FC trapped flux lines go to zero at the same, finite, magnetic field value.

\subsection{Sample in DAC at 2.2 GPa}

\begin{figure}
\begin{center}
\includegraphics[angle=0,width=150mm]{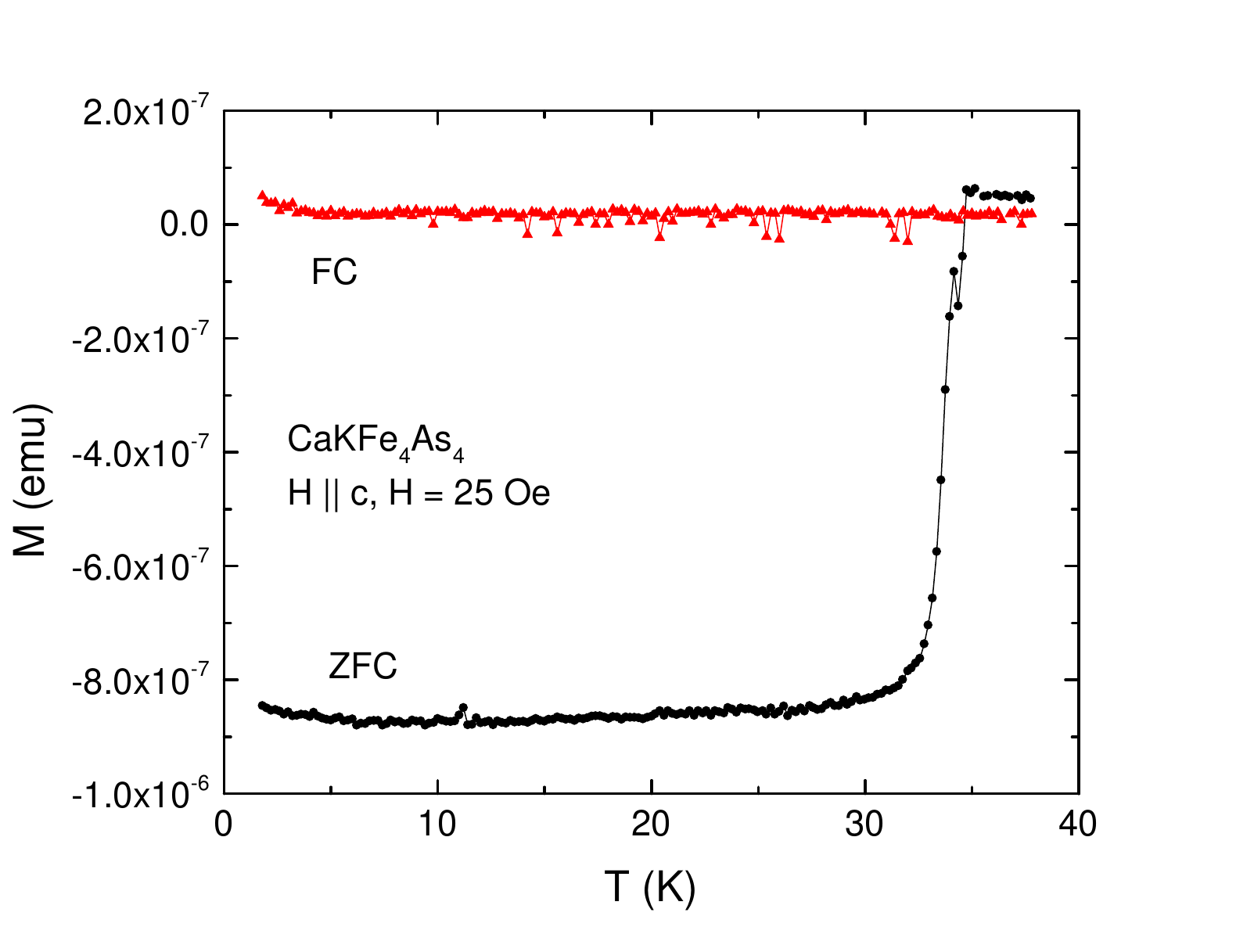}
\end{center}
\caption{(color online) Temperature dependent zero-field-cooled (ZFC) and field-cooled (FC) magnetization, $M(T)$ of of an $\sim 0.5 \mu$g (see text for details) sample of CaKFe$_4$As$_4$ measured in 25 Oe magnetic field applied along the $c$-axis.} \label{F1} 
\end{figure}

\begin{figure}
\begin{center}
\includegraphics[angle=0,width=150mm]{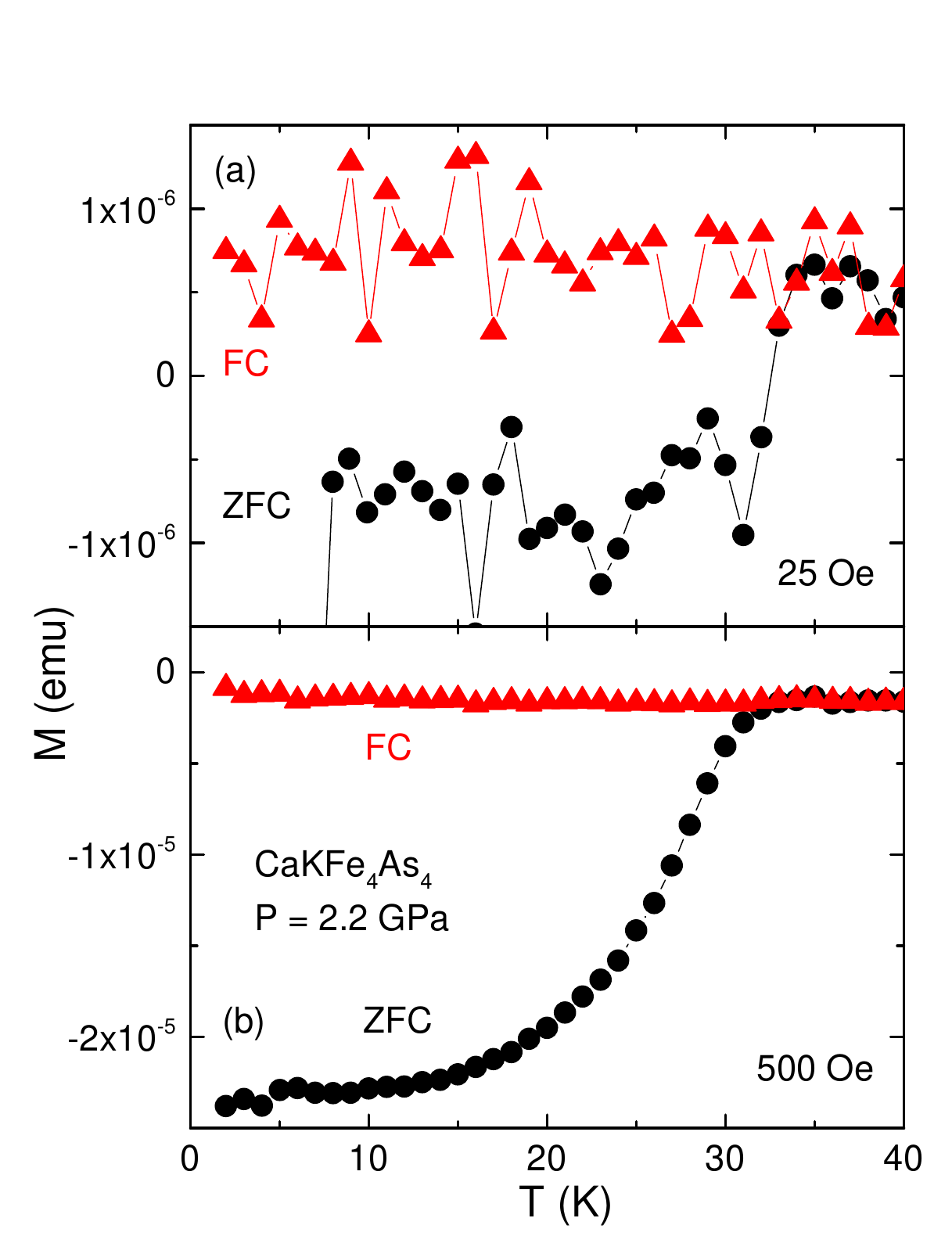}
\end{center}
\caption{(color online) Temperature dependent zero-field-cooled (ZFC) and field-cooled (FC) magnetization, $M(T)$ of CaKFe$_4$As$_4$ in DAC under pressure of 2.2 GPa measured in (a) 25 Oe and (b) 500 Oe magnetic field applied along the $c$-axis.} \label{F4} 
\end{figure}

The same sample of CaKFe$_4$As$_4$ was subsequently placed in a DAC, the pressure was increased to 2.2 GPa and magnetization measurements were performed. ZFC and FC $M(T)$ data in 25 Oe and 500 Oe applied field are shown in Fig. \ref{F4}. The superconducting transition is seen in both sets of measurements with $T_c \approx 33.8$~K in the 25 Oe ZFC data set. This decrease of $T_c$ under pressure is comparable with the literature. \cite{kal17a} It has to be mentioned that the 25 Oe data in Fig. \ref{F4} appear to be noisy and the signal in the normal state seems to be measurably larger than at ambient pressure (Fig. \ref{F1}). The possible reasons include rather small signal from the sample on top of fairly large background (see Appendix B for the examples of the subtraction and Appendix A for discussion of the low temperature background associated with the commercial DAC that was used), possibly slightly different remnant field in the superconducting magnet in in background and sample measurements among others. The $M(T)$ data at 500 Oe [shown in Fig. \ref{F4}(b)] are less noisy due to a larger signal (see Appendix C for more on this), the apparent transition seen in ZFC data is broader, in agreement with larger applied field and thin-plate-like geometry of the sample with magnetic field being perpendicular to the plate.  

\begin{figure}
\begin{center}
\includegraphics[angle=0,width=120mm]{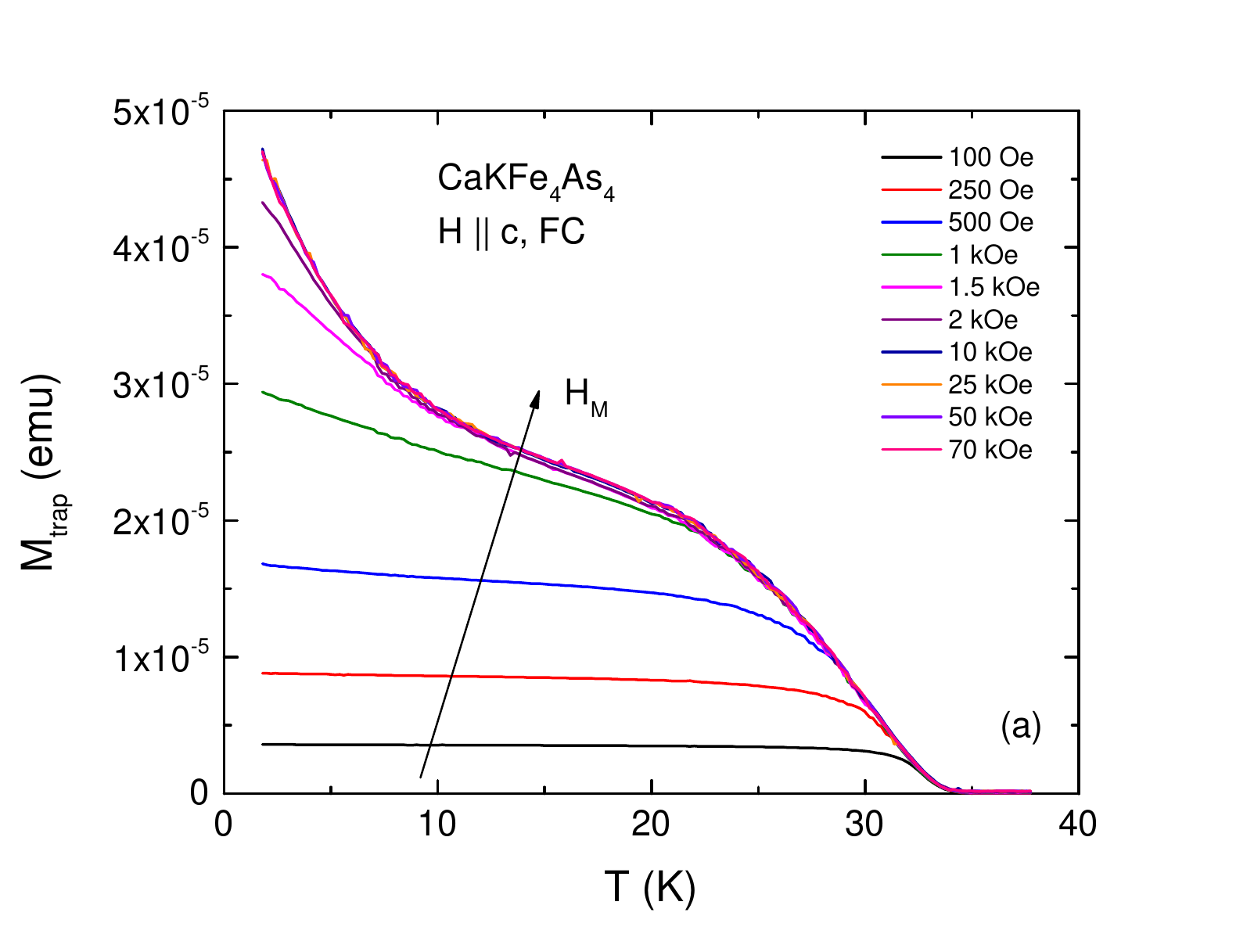}
\includegraphics[angle=0,width=120mm]{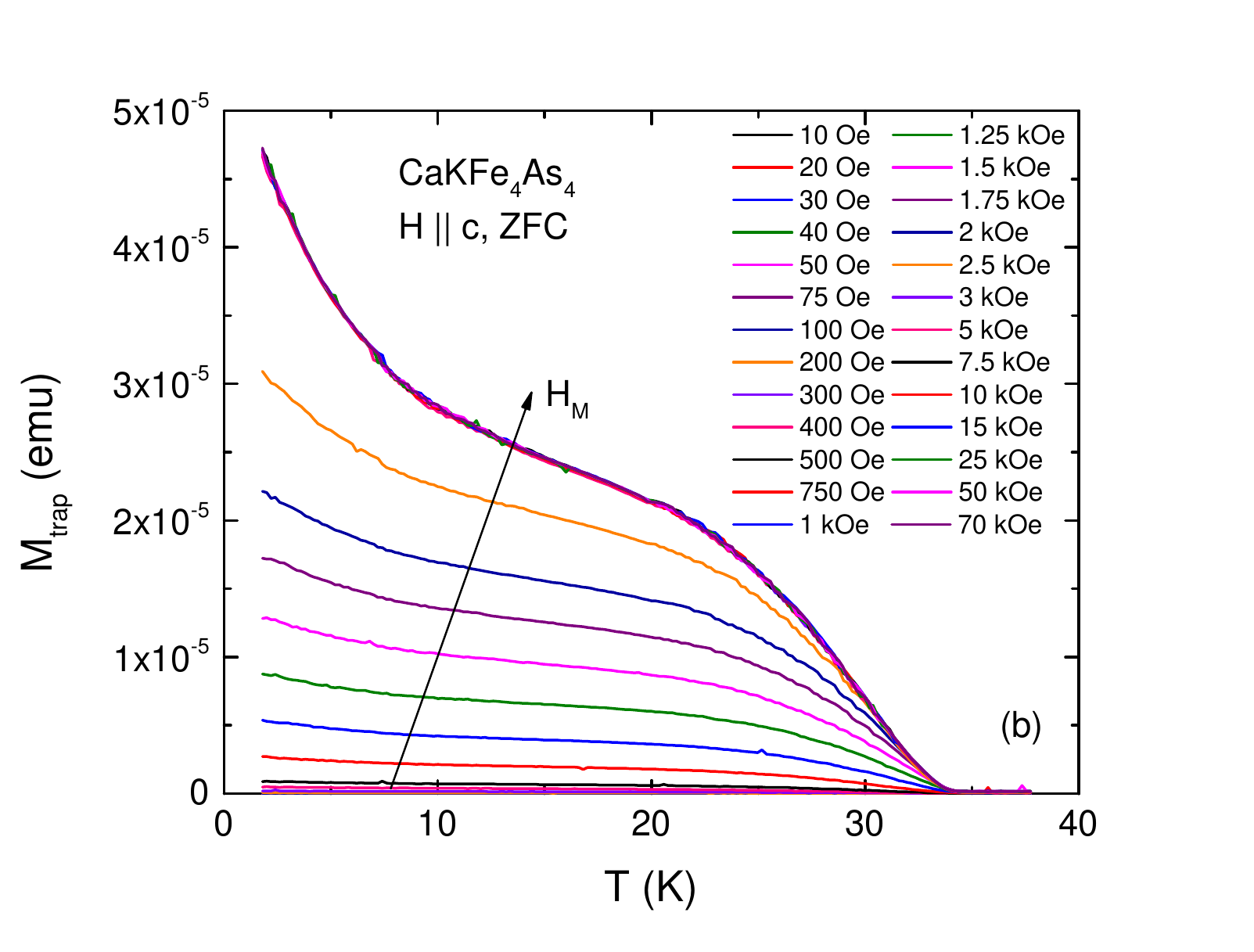}
\end{center}
\caption{(color online) Temperature dependent trapped flux magnetization of CaKFe$_4$As$_4$ measured in $H = 0$ using (a) FC and (b) ZFC protocols. Legends show target magnetic fields, $H_M$, in FC  and ZFC experiments. Arrows point the direction of increase of $H_M$.} \label{F2} 
\end{figure}

\begin{figure}
\begin{center}
\includegraphics[angle=0,width=120mm]{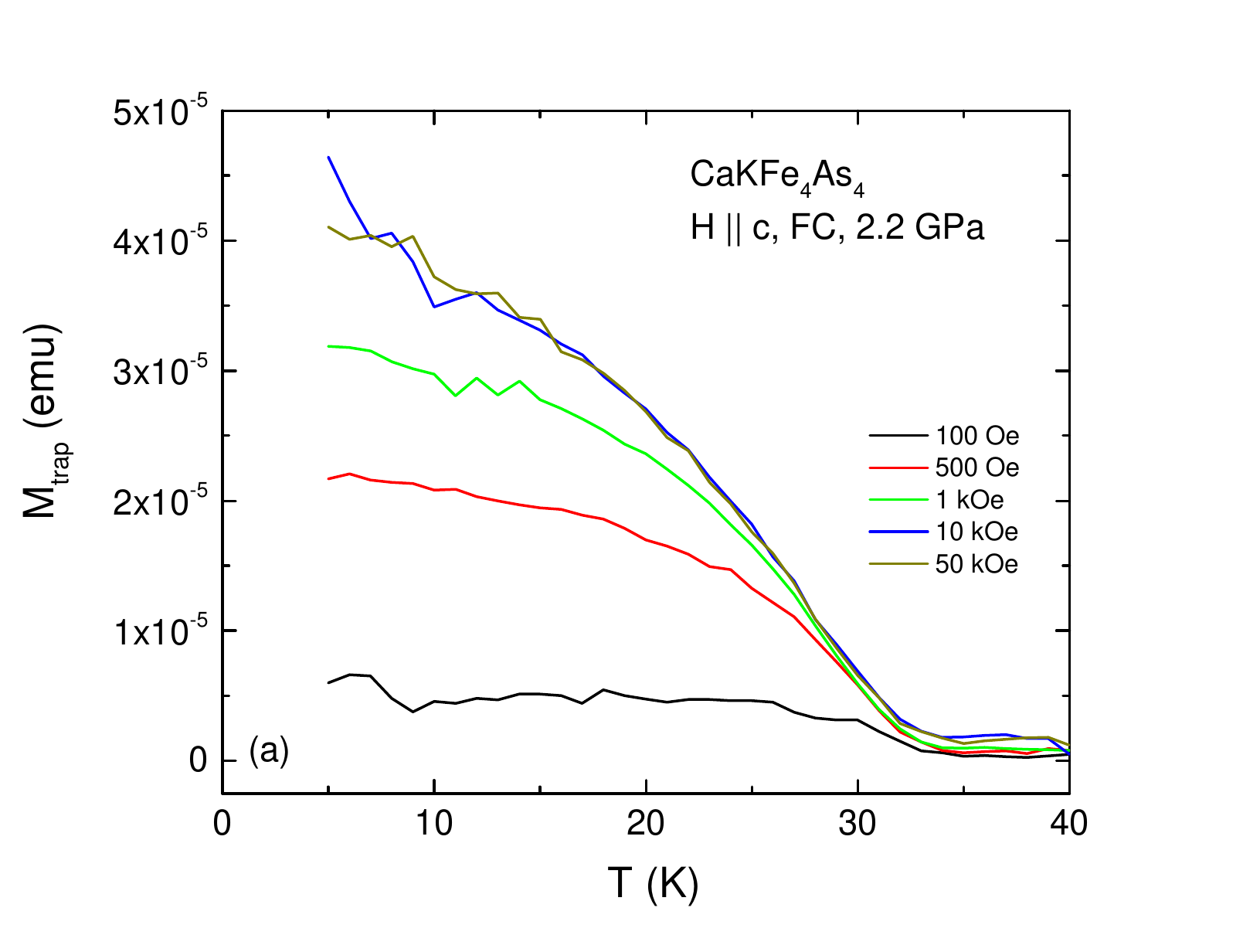}
\includegraphics[angle=0,width=120mm]{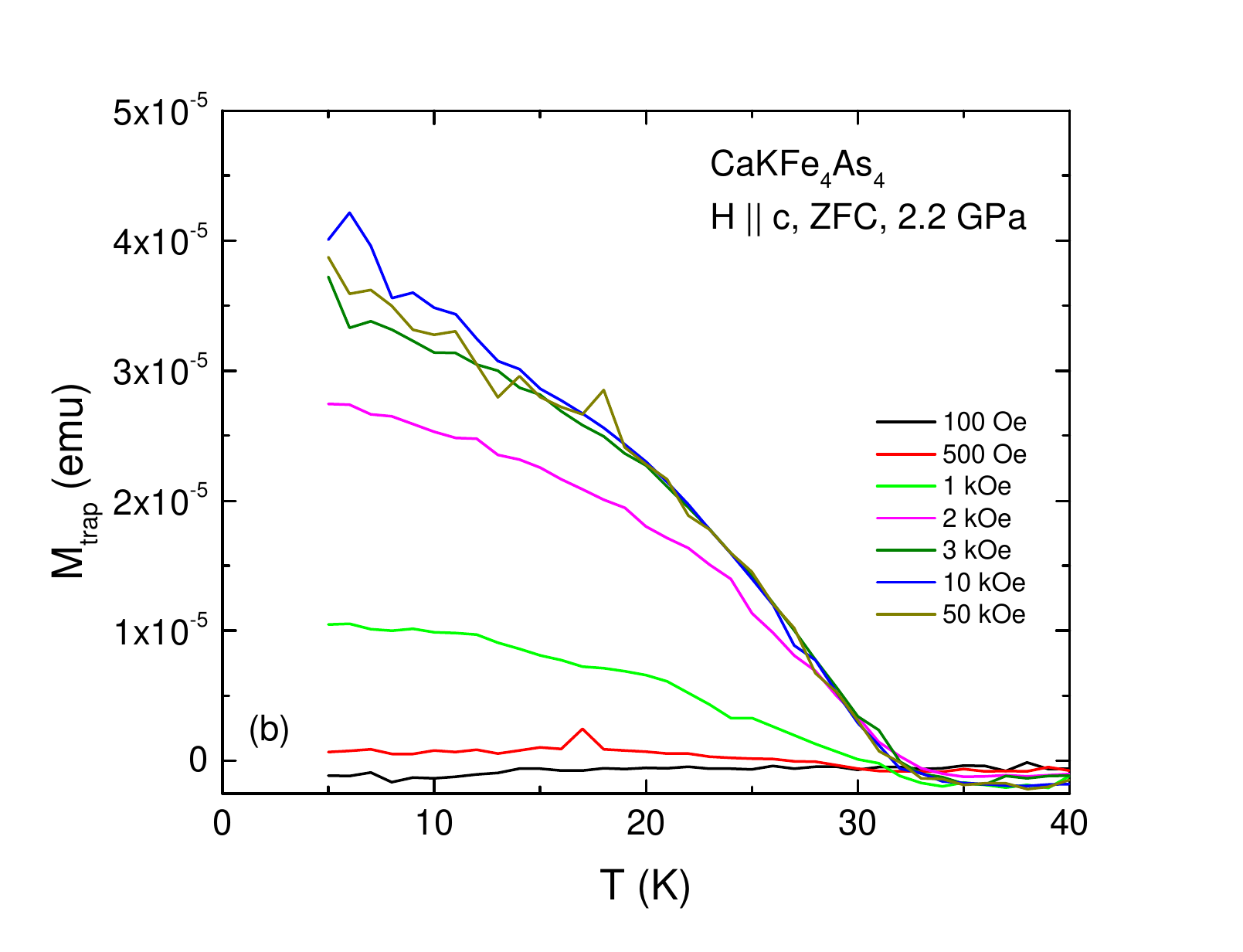}
\end{center}
\caption{(color online) Temperature dependent trapped flux magnetization of CaKFe$_4$As$_4$ measured in $H = 0$ using (a) FC and (b) ZFC protocols in DAC under pressure of 2.2 GPa. Legends show target magnetic fields, $H_M$ in FC  and ZFC experiments.} \label{F5} 
\end{figure}

\begin{figure}
\begin{center}
\includegraphics[angle=0,width=120mm]{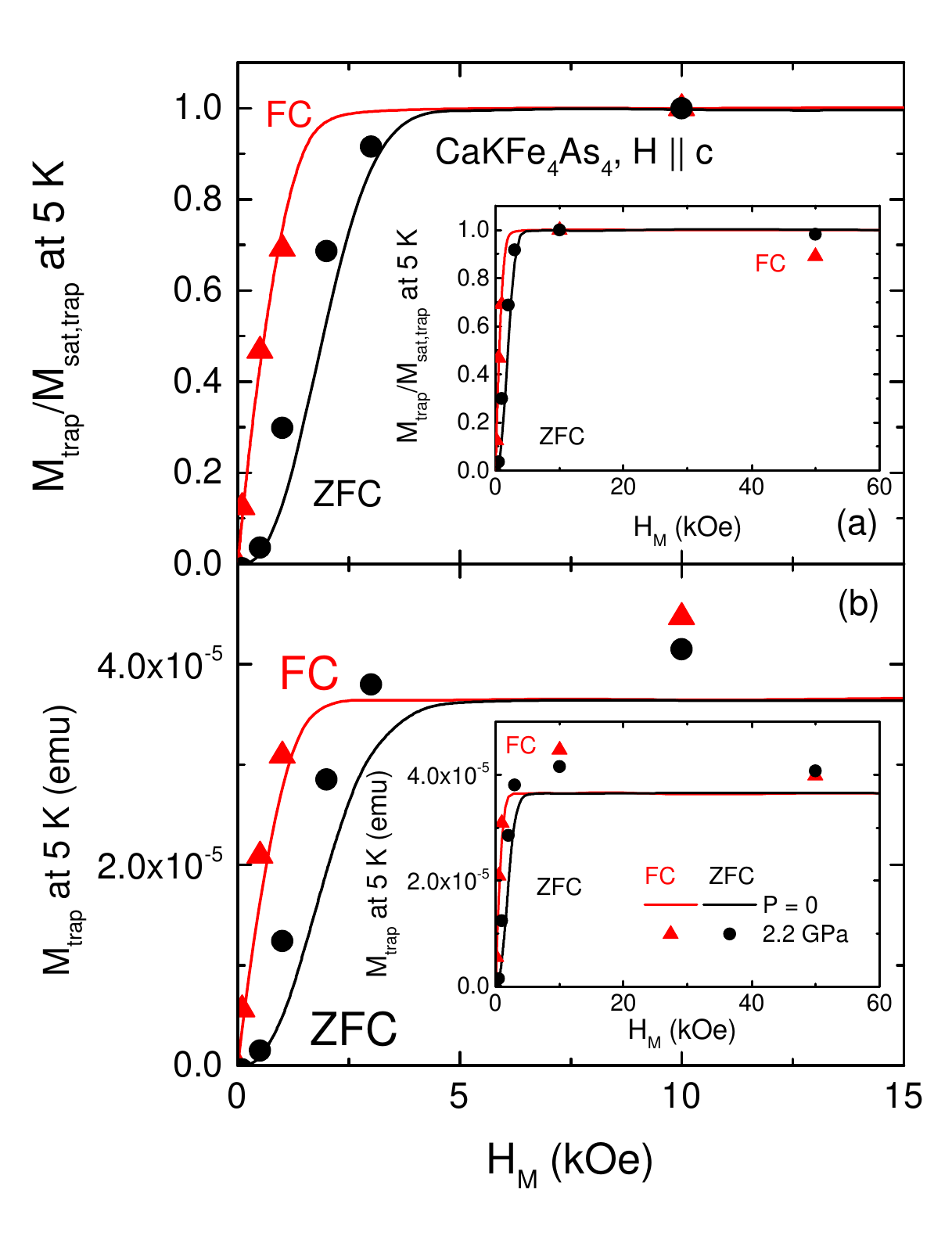}
\end{center}
\caption{(color online) (a) Normalized to values at $H_M = 10$~kOe trapped field magnetization at $T = 5$~K as a function of target field $H_M$ in the ZFC and FC experiments. Black -ZFC, red FC protocols. Lines - $P = 0$, filled symbols - measurements in DAC at 2.2~GPa. The data shown for $H_M \leq 15$~kOe. Panel (b) shows the same data without normalization. Insets show the corresponding data up to $H_M = 60$~kOe.} \label{FA5} 
\end{figure}

Data for temperature dependent magnetization associated with the trapped flux for the sample in DAC are shown in Fig. \ref{F5}. They are similar to those obtained at the ambient pressure (Fig. \ref{F2}) but with somewhat higher apparent noise levels . There is also an apparent difference in the higher temperature, normal state zeros for the two data sets.  This is associated with inherent uncertainty of inferring a zero signal by subtracting two large numbers. A comparison of trapped field magnetization at 5~K vs target magnetic field, $H_M$ under pressure and at ambient pressure  is shown in Fig. \ref{FA5}. Given that the comparison is for 5~K, the data in Fig. \ref{FA5} for the ambient pressure results are slightly different from the 1.8~K data plotted in Fig. \ref{F3} (above).  The normalized data look remarkably similar. The same data in the absolute values (shown in the inset) show some difference, in particular in saturated state, at 10~kOe. For the pressure data, this discrepancy is likely a combination of some real increase of the trappped field magnetization (i.e. critical current density) under pressure combined with increased error bars of the measurements.

\section{Discussion and summary}

\begin{figure}
\begin{center}
\includegraphics[angle=0,width=150mm]{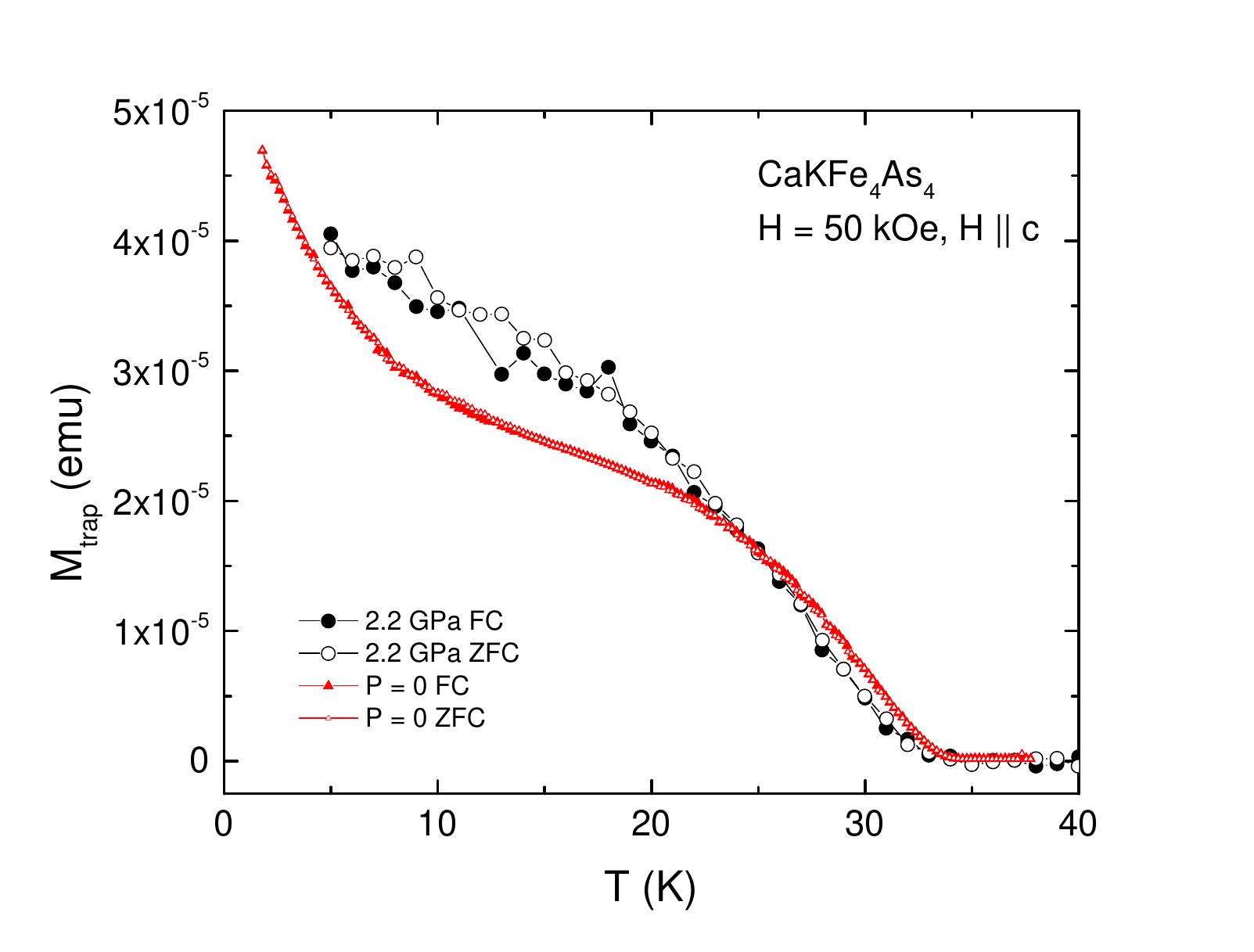}
\end{center}
\caption{(color online) Comparison of  temperature dependent trapped flux magnetization of CaKFe$_4$As$_4$ at ambient and 2.2~GPa pressure measured in nominal $H = 0$ after $H_M = 50$~kOe using FC and ZFC protocols.  Data at 2.2~GPa were shifted vertically, so that they are zero in 35 - 40~K temperatures range.} \label{F6} 
\end{figure}

To have a more direct comparison of the trapped field related magnetization, in Fig. \ref{F6} we plot  $H_M = 50$~kOe FC and ZFC magnetization data taken at ambient pressure and in DAC under 2.2~GPa. At $H_M = 50$~kOe we are well in the saturation region for the trapped field. For this comparison plot,  2.2~GPa data were shifted vertically so that $M_{trap} \approx 0$ in 35 - 40 K temperature region. These shifts, $\sim 1 \times 10^{-6}$~emu in the absolute values, are probably a consequence of several experimental and data analysis issues: accuracy of subtraction of background, reproducibility of the axial and radial position of the DAC as well as reproducibility of the remnant field in the superconducting magnet in background and sample measurements and others. Additional experimental effort is needed to further understand and possibly reduce these systematic deviations.  With the foregoing all stated, we can make several observations: as expected, FC and ZFC  data are laying on top of each other; the data show that $T_c$ is slightly suppressed at 2.2 GPa; whereas at higher temperatures $M_{trap}(T)$ data at ambient and 2.2 GPa pressure are similar, there is a notable difference below $\sim 25$~K. This difference could be intrinsic, as saturated trapped field magnetization is proportional to the self-field critical current density. \cite{bud23a,min23a}  which could be affected by pressure. \cite{bud93a,jun15a} It is possible that the already anomalous (at ambient pressure) temperature dependence of critical current density (which is possibly related to CaFe$_2$As$_2$ intergrowths, as suggested in Ref. \cite{ish19a}) changes even further as the pinning landscape evolves under pressure. Additionally, part of this difference might be related to a possible deformation of the sample in a DAC under pressure. 

Now we need to turn our attention to the important difference in protocols for trapped flux magnetization measurements between this work and Ref. \cite{min23a}.    The trapped flux magnetization measurements are performed in {\it nominal} zero applied field. As it was discussed in detail in Ref. \cite{bud23a}, in reality there is a remnant magnetic field in the superconducting magnet of a  SQUID magnetometer used for the measurements which depends on the superconducting magnet design and geometry as well as the history of magnetic fields applied prior to the measurement. In 70 kOe MPMS3 magnet the remnant field can be as high as 25 Oe. \cite{bud23a} Then the trapped flux magnetization signal from the sample would be entangled with the signal from the DAC in remnant field. The signal from the DAC depends on the details if the design and geometry of the DAC and materials used in the DAC. Almax - easyLab Mcell Ultra DAC \cite{mcell} used in this work is made out of a Cu-Be alloy, other parts used are diamond anvils, tungsten gasket, plus very small amounts of pressure medium and  epoxy used to glue the anvils. Of note on the geometry is the presence of several access ports in the vicinity of the sample position, that would contribute to the background signal. The DAC in Ref. \cite{min23a} is made mainly from Cu - 3 wt.\% Ti alloy, Cu - Be was used for a small piston and an anvil seat, other parts are diamond anvils, rhenium gasket and small amounts of epoxy. \cite{dro15a,ere22a} The main materials used in the construction of these two DACs, Cu - Be and Cu - Ti alloys, have significantly different magnetic susceptibility (see Appendix A) with Cu - Ti susceptibility being  4 - 6 times lower than that of Cu - Be, and staying below $1 \times 10^{-7}$~emu/g in the whole temperature range. In addition, the measurements in this work cover $T \leq 40$~K temperature range, where Cu - Be has a distinct upturn in susceptibility. A combination of these factors explains the need for point-by-point background subtaction, even for trapped flux magnetization measurements, for the commercial DAC.

One of the possible modifications of the protocol of the trapped flux magnetization measurements, that could help to further mitigate cell background concerns further,  would be to perform the measurements not in nominal $H = 0$, but in a finite $H = - H_{rem}$ magnetic field ($H_{rem}$ is a remnant magnetic field in a superconducting magnet that depends of the magnet history). Two possible ways to perform such calibration were presented in the Appendix of Ref. \cite{bud23a}. The examples of the (partial) offsets of the remnant field by measurements in finite applied field were shown in the same publication having MgB$_2$ crystal at ambient pressure as an example.

To summarize, the data presented in this work show that it is possible to do reasonable magnetic measurements, including magnetization related to the trapped flux,  on very small, similar in size to those in Ref. \cite{min23a}, superconducting samples in DAC. Moreover, with an appropriate care (point-by-point background subtraction), a commercial, Cu - Be DAC can be suitable for such measurements.

\ack

This work was supported by the U.S. Department of Energy, Office of Science, Basic Energy Sciences, Materials Sciences and Engineering Division. Ames National Laboratory is operated for the U.S. Department of Energy by Iowa State University under Contract No. DE- AC02-07CH11358. PCC thanks Shanti Deemyad for remarks that helped motivate this work.

\clearpage

\clearpage

\clearpage

\clearpage

\clearpage

\clearpage

\clearpage

\clearpage 

\appendix

\clearpage
\section{Magnetic properties of DAC construction materials}

\begin{figure}
\begin{center}
\includegraphics[angle=0,width=150mm]{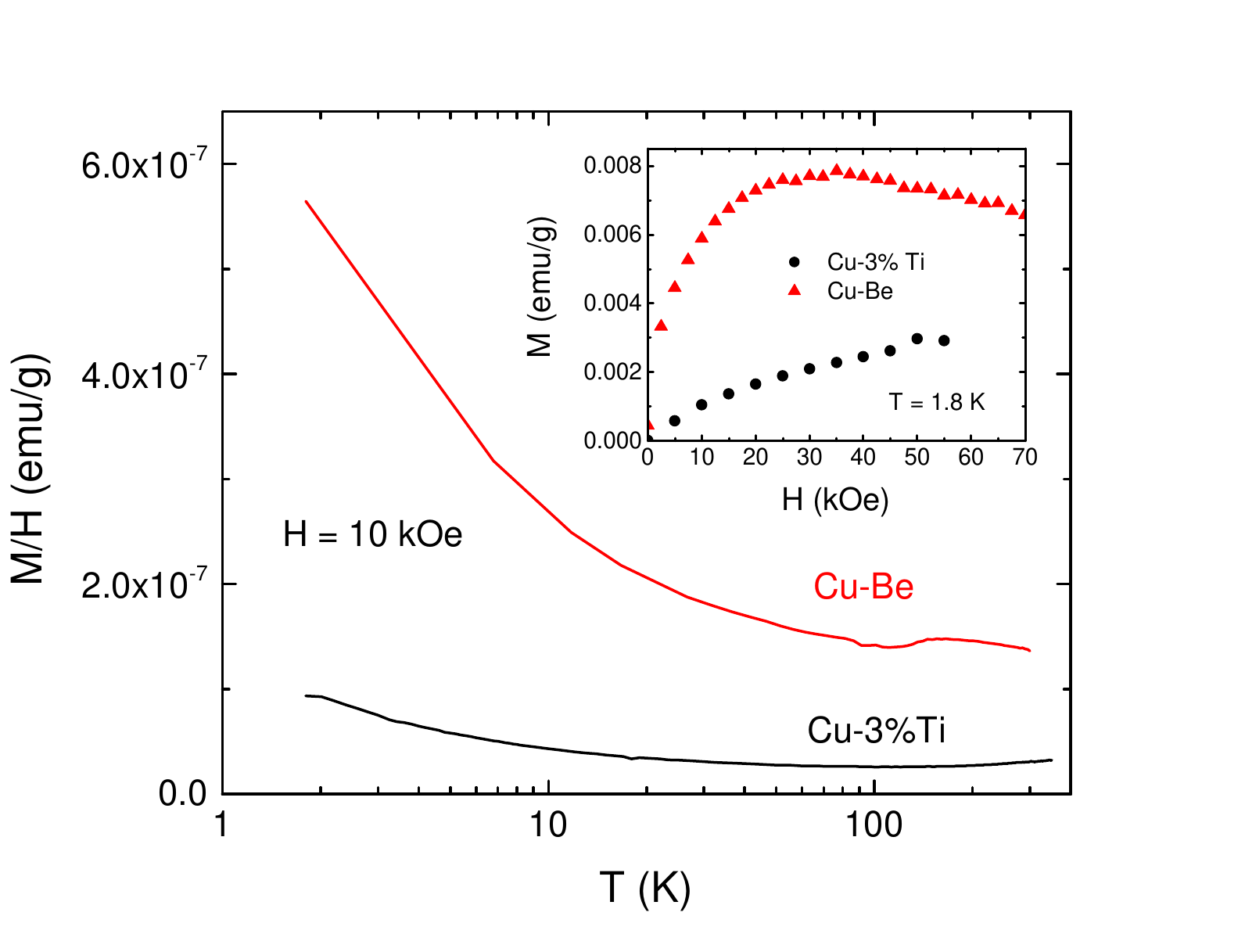}
\end{center}
\caption{(color online) Temperature-dependent magnetic susceptibility, $M/H(T)$, of heat treated Cu - 3 wt.\% Ti alloy and Cu - Be alloy measured in 10~kOe applied field. Inset: field - dependent magnetization measured at $T = 1.8$~K. Note that for Cu - Be alloy measurements a set screw from our Almax - easyLab Mcell Ultra was taken with expectation that it reflects the magnetic properties of the particular Cu - Be batch used to manufacture the DAC.} \label{FA2} 
\end{figure}

Figure \ref{FA2} presents the results of magnetic measurements on Cu - 3 wt.\% Ti alloy and Cu - Be alloy on semi-log scale. Be - Cu commercial alloy contains $\sim 97.9$~wt. \% of Cu, $\sim 1.9$~wt. \% of Be and $\sim 0.2$~wt. \% of Co. Commercial (small batch) Cu - 3 wt.\% Ti alloy nominally does not include magnetic components, but the data show, smaller than in Be - Cu, low temperature, upturn that probably is associated with minor transition metal impurities in the startng materials.   Pure elemental Cu is diamagnetic. Consequently, the $M(H)$ of the Be - Cu alloy at the base temperature is a combination of a Brillouin - like increase and saturation of  $M(H)$ associated with Co contribution followed by a slow decrease due to diamagnetism of Cu. $M(H)$ of the Cu - 3 wt.\% Ti alloy apparently did not reach a clear saturation below 55~kOe. Note that in the whole temperature range magnetic susceptibility, $M/H$, of Cu - Be is significantly, 4 - 6 times, higher than that of Cu - 3 wt.\% Ti, and the low temperature upturn in Cu - Be is more pronounced.

\clearpage

\section{Examples of background subtraction}

\begin{figure}
\begin{center}
\includegraphics[angle=0,width=80mm]{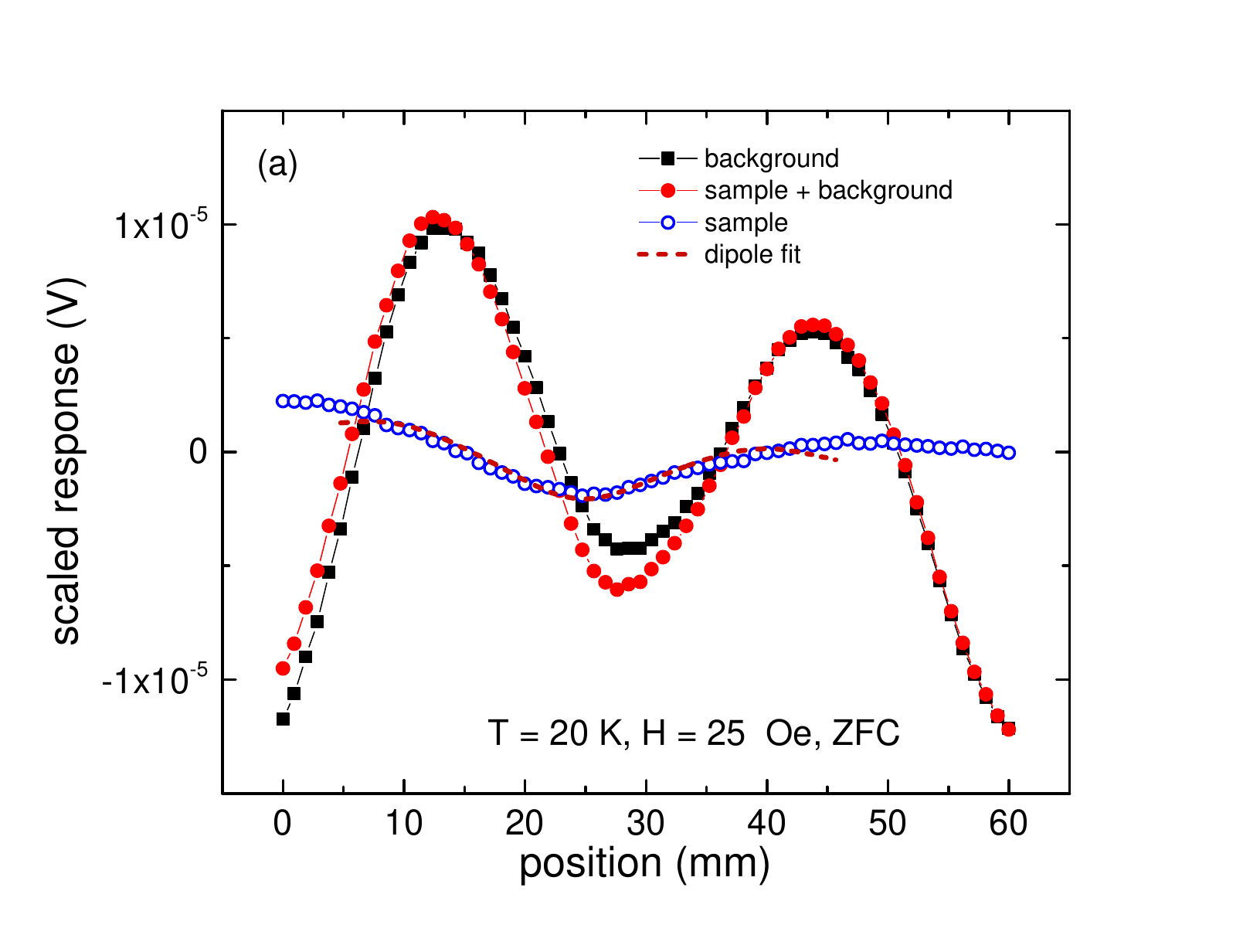}
\includegraphics[angle=0,width=80mm]{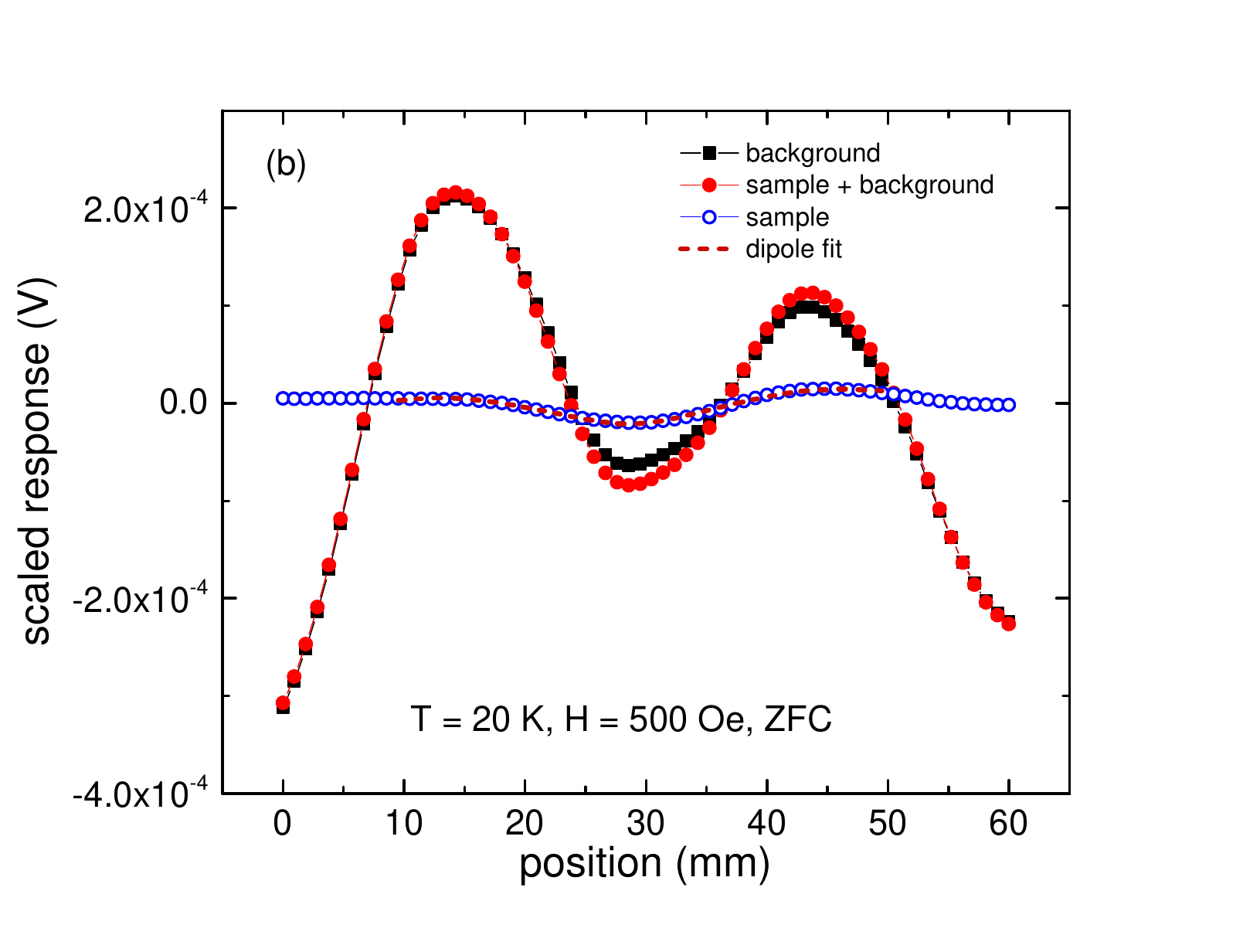}
\includegraphics[angle=0,width=80mm]{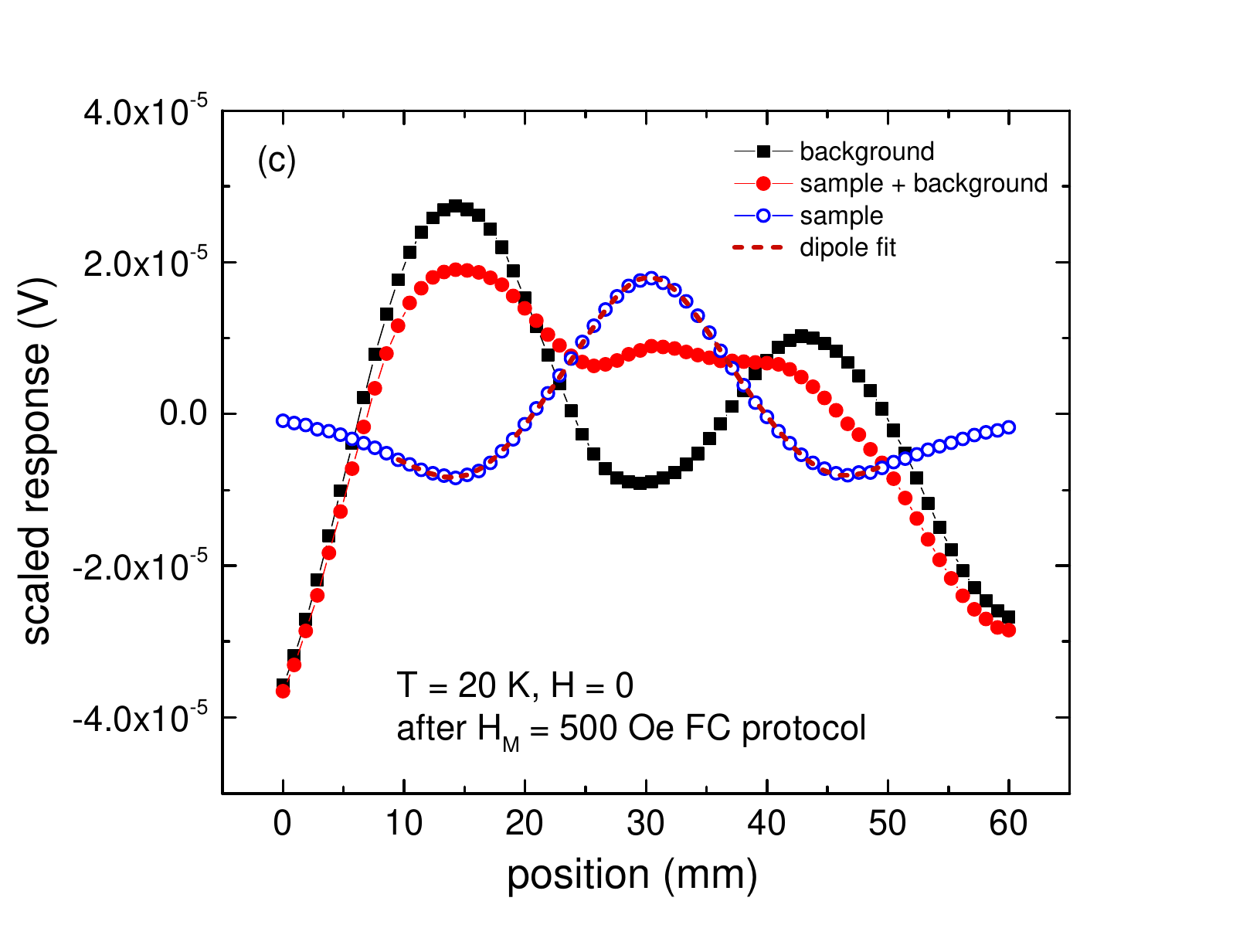}
\end{center}
\caption{(color online) Examples of pont-by-point background subtraction in measurements of CaKFe$_4$As$_4$ under 2.2 GPa pressure in a commercial DAC. All examples are for $T = 20$~K. (a) $M(T)$ ZFC measurements, $H = 25$~Oe; (b) $M(T)$ ZFC measurements, $H = 500$~Oe; (c) nominal $H = 0$ measurements of trapped flux magnetization using FC protocol with the target field $H_M = 500$~Oe. } \label{FA1} 
\end{figure}

Figure \ref{FA1} shows three examples of background subtraction for measurements of CaKFe$_4$As$_4$ crystal in DAC. For more details of the procedure refer to Ref. \cite{coa20a}.

\clearpage

\section{Noise in FC magnetization measurements under pressure}

\begin{figure}
\begin{center}
\includegraphics[angle=0,width=120mm]{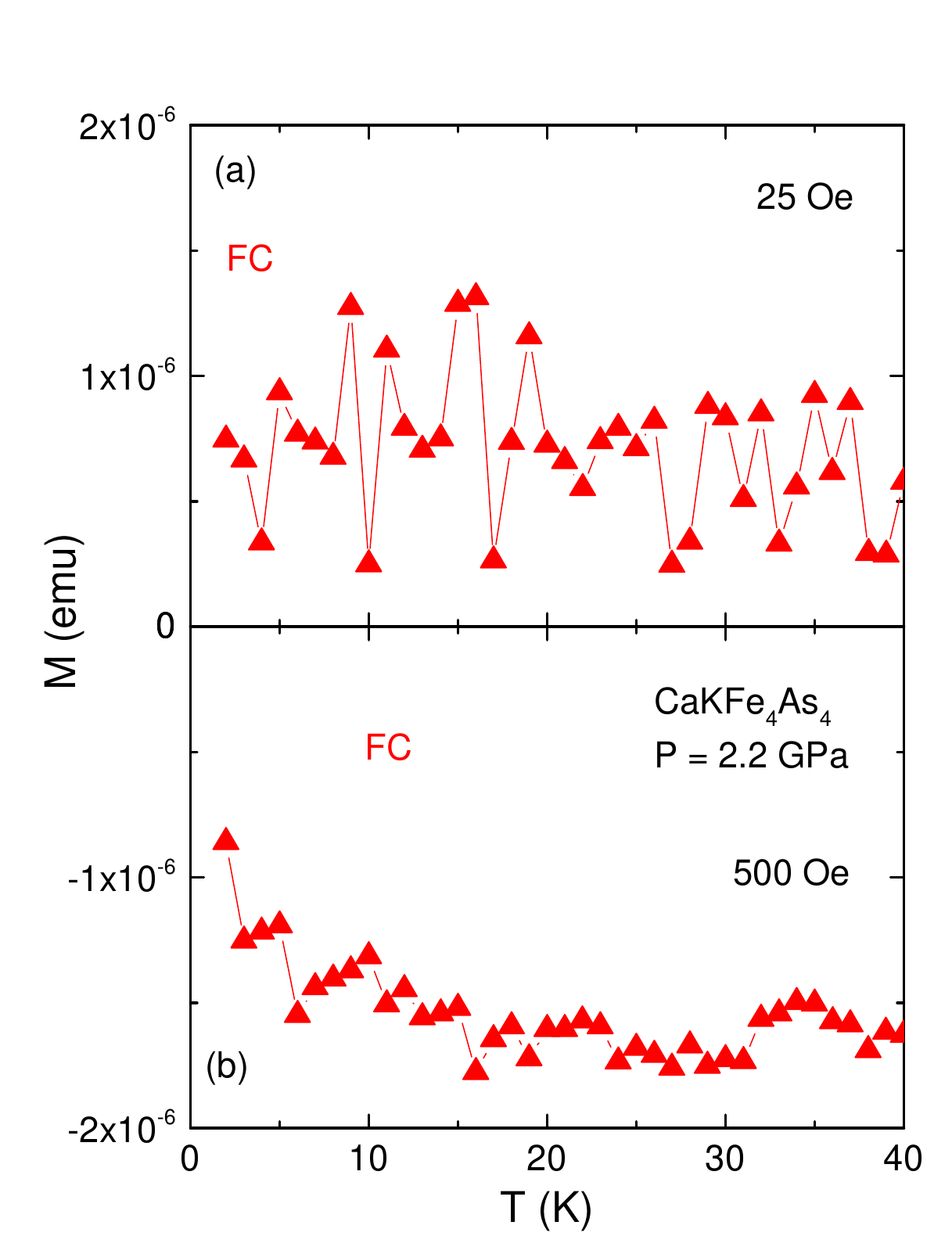}
\end{center}
\caption{(color online)  FC $M(T)$ measurements of CaKFe$_4$As$_4$ crystal in DAC in two different applied fields, (a) 25~Oe and (b) 500~Oe from Fig. \ref{F4} plotted on the Y - axis scales that span the same $2 \times 10^{-6}$~emu. } \label{FA3} 
\end{figure}

Figure \ref{FA3} shows FC $M(T)$ measurements of CaKFe$_4$As$_4$ crystal in DAC in two different applied fields, 25~Oe and 500~Oe presented above in Fig. \ref{F4} on the Y - axis scales that span the same $2 \times 10^{-6}$~emu. The random point-to point noise appears to be smaller for 500~Oe data, but some drift is present in the data. It is possible that an inherent uncertainty caused by large background subtraction and  limited resolution of the MPMS-classic contributes to this noise difference.

It is noteworthy, that a small feature corresponding to $T_c$ apparently can be seen in the 500~Oe FC data.

\clearpage

\end{document}